\documentclass[a4paper,12pt]{article}
\usepackage{amsmath,amssymb,cite}
\usepackage[english]{babel}
\usepackage{graphicx}
\setkeys{Gin}{keepaspectratio}
\usepackage[hyperindex=false,hyperfootnotes=false,colorlinks=true,linkcolor=blue,citecolor=blue,urlcolor=blue,pdftitle={Linear and Nonlinear Evolution and Diffusion Layer Selection in Electrokinetic Instability},pdfauthor={E. A. Demekhin, V. S. Shelistov and S. V. Polyanskikh}]{hyperref}
\newcommand{\header}[1]{\pagebreak[3]\textit{#1}}
\providecommand{\url}[1]{#1}	% in case hyperref is commented out

\title{Linear and Nonlinear Evolution and Diffusion Layer Selection in Electrokinetic Instability}

\author{E. A. Demekhin $^{[1]}$, V. S. Shelistov and S. V. Polyanskikh\\
Dept. of Computation Mathematics and Computer Science\\
Kuban State University\\
Krasnodar, 350040, Russian Federation.}

\begin{document}

\setlength{\baselineskip}{18pt} \maketitle
\begin{center}
Abstract
\end{center}
In the present work fournontrivial stages of electrokinetic instability are
identified by direct numerical simulation (DNS) of the full
Nernst--Planck--Poisson--Stokes (NPPS) system:
i) The stage of the influence of the initial conditions (milliseconds);
ii) 1D self-similar evolution (milliseconds--seconds);
iii) The primary instability of the self-similar solution (seconds);
iv) The nonlinear stage  with secondary instabilities.
The self-similar character of evolution at intermediately large times is
confirmed. Rubinstein and Zaltzman instability and noise-driven nonlinear
evolution to over-limiting regimes in ion-exchange membranes are numerically
simulated and compared with theoretical and experimental predictions. The
primary instability which happens during this stage is found to arrest
self-similar growth of the diffusion layer and specifies its characteristic
length as was first experimentally predicted by Yossifon and Chang
(PRL \textbf{101}, 254501 (2008)). A novel principle for the characteristic wave
number selection from the broadbanded initial noise is established.

\footnotetext[1] {E-mail adress: edemekhi@gmail.com}

\setlength{\baselineskip}{24pt}
\renewcommand{\thesection}{\Roman{section}{.}}
\renewcommand{\thesubsection}{\Alph{subsection}{.}}

\header{Introduction.}
Pattern formation in dissipative systems has a long and intriguing history in a
number of disciplines. It is generally associated with nonlinear effects which
lead to qualitatively new phenomena that do not occur in linear systems
\cite{CrHog}. Problems of electro-kinetics has recently attracted a great deal
of attention due to the rapid development of micro-, nano- and biotechnologies.
A curious and fascinating electrokinetic instability and pattern formation in an
electrolyte solution between semi-selective ion-exchange membranes under a DC
potential drop was theoretically predicted in \cite{Zlk,Rub13} and
experimentally confirmed in \cite{Rub23}, including direct experimental proof
\cite{RubRub1,KmHn,YCh}. Both theory and experiments show that these
instabilities are reminiscent of Rayleigh--Benard convection and
Benard--Marangoni thermo-convection, but are much more complicated from both the
physical and mathematical points of view.

In the present work  for the first time nontrivial stages of the noise-driven
electrokinetic instability are obtained by direct numerical simulation (DNS) of
the full Nernst--Planck--Poisson--Stokes (NPPS) system. Small-amplitude initial
room disturbances have the following stages of evolution:
i) A transient stage of the influence of the initial conditions,
$\tilde{t}=O(\tilde{\lambda}_D^2/ 4 \tilde{D})$ (milliseconds);
ii) A 1D self-similar stage of evolution, $\tilde{\lambda}_D^2/4 \tilde{D} \ll
\tilde{t} \ll \tilde{L}^2/4 \tilde{D}$ (milliseconds--seconds);
iii) A stage in which instability becomes manifest and arrests any self-similar
growth and selects a diffusion layer characteristic length,
$\tilde{t}=O(\tilde{L}^2/4\tilde{D}$) (seconds);
iv) A nonlinear stage  with secondary instabilities, $\tilde{t} \gg
\tilde{L}^2/4 \tilde{D}$.

\header{Formulation.}
A binary electrolyte between perfect semi-selective ion-exchange membranes is
considered. Notations with tilde are used for the dimensional variables, as
opposed to their  dimensionless counterparts without tilde. The diffusivities of
cations and anions are assumed to be equal, $\tilde{D}^+=\tilde{D}^-=\tilde{D}$.
The following characteristic values are taken to make the system dimensionless:
distance between the membranes, $\tilde{L}$; dynamic viscosity, $\tilde{\mu}$;
thermodynamic potential, $\tilde{\Phi}_0= \tilde{R}\tilde{T}/\tilde{F}$; bulk
ion concentration at initial time, $\tilde{c}_0$. Then, the characteristic time
and characteristic velocity are scaled, respectively, by $\tilde{L}^2/\tilde{D}$
and $\tilde{D}/\tilde{L}$. Here $\tilde{F}$ is Faraday's constant; $\tilde{R}$
is the universal gas constant; $\tilde{T}$ is Kelvin temperature; $\tilde{\mu}$
is the dynamic viscosity of the fluid; $\tilde{\varepsilon}$ is the permittivity
of the medium; and the dimensional Debye length is taken as
$\tilde{\lambda}_D=\sqrt{\tilde{\varepsilon} \tilde{R} \tilde{T}/
\tilde{F}^2\tilde{c}_0}$.

Electro-convection is then described by the transport equations for the
concentrations for positive and negative ions, ${c}^+,\ {c}^-$, the Poisson
equation for the electric potential, $\Phi$, and the Stokes equation for a
creeping flow, presented in dimensionless form:
\begin{equation}\label{eq1Dim}
\begin{array}{cc}
\displaystyle \frac{\partial c^\pm}{\partial t} + \mathbf{U}\cdot{\nabla}c^\pm
 = \pm\nabla\cdot\left(c^\pm\nabla\Phi\right) + \nabla^2 c^\pm,\\
\nu^2\nabla^2\Phi = c^- - c^+,
 \quad \nabla P - \nabla^2\mathbf{U} = \kappa\nabla^2\Phi\nabla\Phi,
 \quad \nabla\cdot\mathbf{U} = 0
\end{array}
\end{equation}
with the boundary conditions at the membrane surfaces, $y=0$ and $1$, a fixed
concentration of positive ions, no flux condition for negative ions, and a fixed
drop of potential and velocity adhesion:
\begin{equation}\label{eq20007Dim}
\displaystyle c^+ = p, \quad -c^-\frac{\partial\Phi}{\partial y}
 + \frac{\partial c^-}{\partial y} = 0, \quad \Phi = 0 \mbox{ for } y=0,
 \quad \Phi = \Delta V \mbox{ for } y = 1, \quad \mathbf{U}=0,
\end{equation}
with the characteristic electric current $j$ at the membrane surface $y=0$,
$j= {c}^+ \partial {\Phi}/\partial {y}+\partial c^+ /\partial {y}$.

The problem is described by three dimensionless parameters: the drop of
potential, $\Delta V$; the dimensionless Debye length, $\nu= \tilde{\lambda}_D
/\tilde{L}$, and the coupling coefficient between hydrodynamics and
electrostatics,
$\kappa=\tilde{\varepsilon}\tilde{\Phi}^2_0/\tilde{\mu}\tilde{D}$ (the last is
fixed for a given electrolyte solution). The dependence on the wall
concentration, $p$, for over-limiting regimes is practically absent \cite{Rub13}
and, hence, $p$ is not included in the mentioned parameters.

The typical bulk concentration of the aqueous  electrolytes varies in the range
$\tilde{c}_0=1\div10^3$~$\mbox{mol}/\mbox{m}^3$; the potential drop is about
$\Delta\tilde{V}=0\div5$~V; the absolute temperature can be taken to be
$\tilde{T}=300^\circ$~K; the diffusivity is about
$\tilde{D}=2\cdot10^{-9}$~$\mbox{m}^2/\mbox{s}$; the distance between the
electrodes $\tilde{L}$ is of the order of $0.5\div1.5$~mm; the concentration
$\tilde{p}$ on the membrane surface must be much larger than $\tilde{c}_0$ and
it is usually taken within the range from $5\tilde{c}_0$ to $10\tilde{c}_0$.
The dimensional Debye layer thickness $\tilde{\lambda}_D$ is varied within the
range from $0.5$ to $15$~nm depending on the concentration $\tilde{c}_0$.

The DNS of the system \eqref{eq1Dim}, without any simplification, is implemented
by applying the Galerkin pseudo-spectral $\tau$-method. A periodic domain along
the membrane surface allows the utilization of a Fourier series, $\exp{(nkx)}$,
in the $x$-direction. Chebyshev polynomials, $T_m(y)$, are applied in the
transverse direction,~$y$. The accumulation of zeros of these polynomials near
the walls allows of properly resolving the thin space charge regions.
Eventually, the physical variables are presented in the form $ R=\sum_{m}
\sum_{n} R_{mn}(t) T_m(y) \exp{(n k x)}$. A special method is developed to
integrate the system in time. The details of the numerical scheme will be
presented elsewhere \cite{DSN}. The basic wave number $k$ is connected with the
membrane length $l$ as $k=2 \pi/l$. Then, $k=1$ and $l=2\pi$ are taken such that
the dimensional length of the membrane $\tilde{l}$ corresponds to the
experimental one in \cite{RubRub1}. The problem is solved for $\nu=10^{-3}$ and
$5\cdot 10^{-4}$, dimensionless drop of potential $\Delta V=0\div200$
($\Delta\tilde{V}=0\div5$~V), and for a typical value of $\kappa$, $\kappa=0.2$;
$p=5$. The results of the full-scale numerical investigation will be accompanied
by a discussion of the main stages of the evolution along with a comparison with
simplified solutions \cite {Dem,KD,DemSh}.

\header{Self-similar stage of evolution.}
Adding initial conditions makes the statement \eqref{eq1Dim}--\eqref{eq20007Dim}
complete. This system is a strongly nonlinear one and a change in the initial
conditions can change the attractor (see, for example, famous experimental work
by Gollub and Swinney \cite{GS} where several attractors, depending on the
initial conditions, were found for flow between two rotating cylinders). Hence
what must be taken are conditions that are natural from the experimental
view-point. From this view-point, when the drop of potential between the
membranes is zero, there are only two thin double-ion layers near the membrane
surfaces and the ion distribution is uniform outside these layers. This
specifies the initial conditions. Then, a small-amplitude white-noise spectrum
should be superimposed on the bulk ion concentrations, $c^+=c^-=1$,
\begin{equation}\label{eq277Dim}
t=0: \quad c^+ - 1 = \mbox{white noise}(x),
 \quad c^- - 1 = \mbox{white noise}(x),
\end{equation}
non-uniform along the membrane. Because of the smallness of the amplitude the
behavior at initial times can be assumed one-dimensional (1D), as if the
right-hand side of \eqref{eq277Dim} were zero, and, as a consequence of this
one-dimensionality, any convection and non-uniformity along $x$ can be discarded
from consideration at initial times: $\mathbf{U}=0$, $\partial/\partial x=0$.

When the potential drop is turned on, an extended space charge (ESC) and
diffusion regions are developed near the bottom membrane. There is an interval
of intermediate times, $\tilde{\lambda}_D^2/4 \tilde{D} \ll \tilde{t} \ll
\tilde{L}^2 / 4 \tilde{D}$, or in dimensionless form, $ \frac{1}{4} \nu^2 \ll t
\ll \frac{1}{4}$, when the problem does not have a static characteristic size:
the double-ion layer is too small and the distance between the membranes is too
large. On dimensional grounds, the only diffusion length which can give a proper
characteristic size that includes time and is, hence, dynamical, is
$\sqrt{4 \tilde{D} \tilde{t}}$ --- and the behavior of the solution has to be
self-similar. These intermediate times vary from milliseconds to seconds (for
details see \cite{Dem,KD,DemSh}).

\begin{figure}[hbtp]
\begin{center}
    \includegraphics[width=7.85cm]{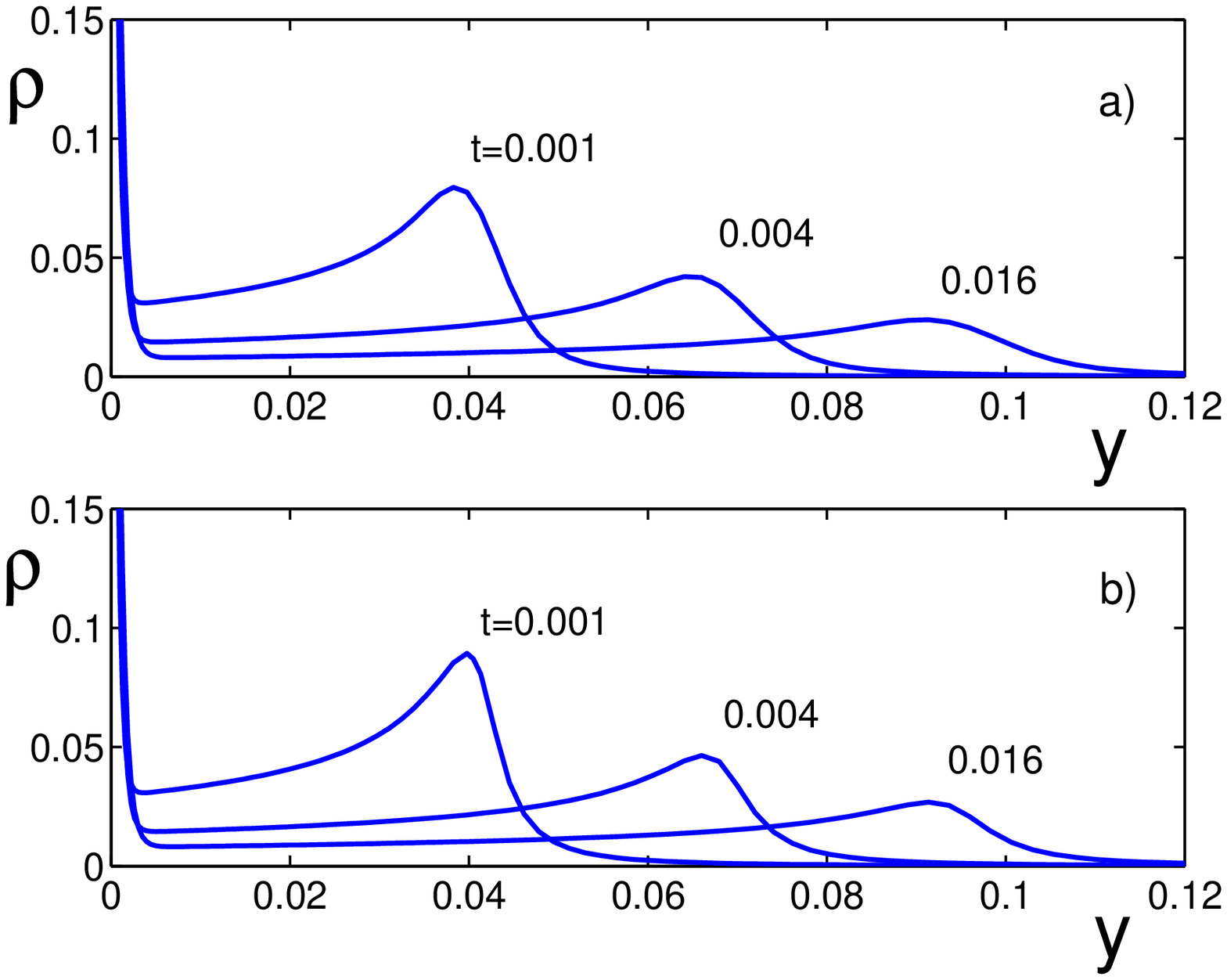}
    \includegraphics[width=7.85cm]{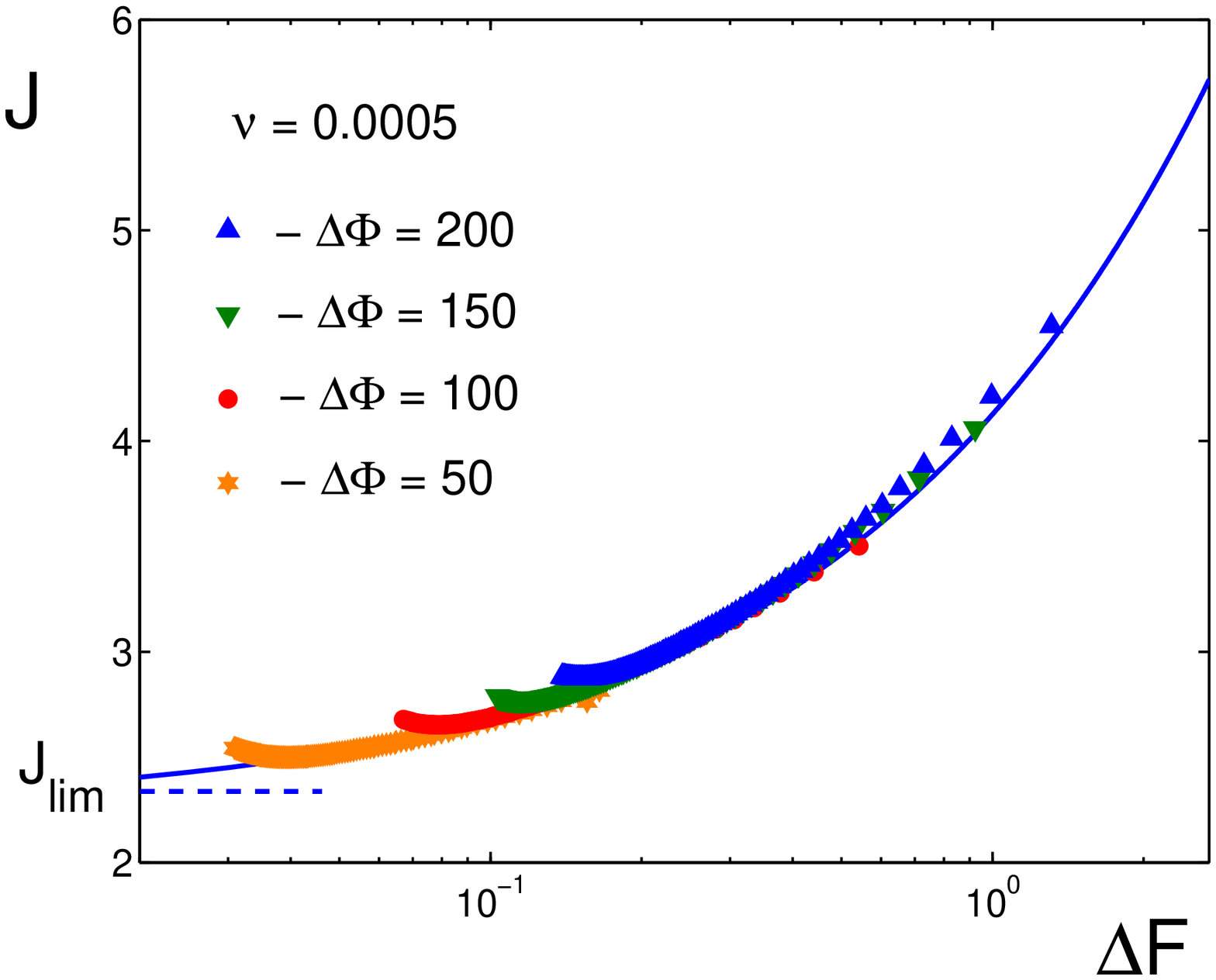}
\end{center}
\caption{(Left) Distribution of the charge density $\rho$ in space for several
moments of time, $\Delta V=100$ and $\nu=\tilde{\lambda_D}/\tilde{L}=0.001$;
(a) --- numerical solution of \eqref{eq1Dim}--\eqref{eq277Dim}, (b) ---
self-similar solution.  The numerical points for $\nu=0.0005$, and several
values of  $\Delta V$ shrink into one self-similar VC curve (shown by the solid
line).}\label{rho}
\end{figure}

The self-similar rescaling of the variables, along with the 1D assumption, turns
the PDE system \eqref{eq1Dim}--\eqref{eq20007Dim} into the ODE system with the
new dimensionless spatial variable $\eta=\tilde{y}/\sqrt{4{\tilde{D}\tilde{t}}}$
and a small parameter
$\varepsilon=\tilde{\lambda}_D/\sqrt{4{\tilde{D}\tilde{t}}}$. A new
dimensionless Debye length $\varepsilon$, electric current $J$ and drop of
potential $\Delta \Phi$ in self-similar variables are connected with the
corresponding old variables, $\nu$, $j$ and $\Delta V$, by the relations
\begin{equation}\label{num100}
 \varepsilon(t)=\frac{\nu}{2 \sqrt{t}},\quad J(t)=2j(t)\sqrt{t}, \quad\Delta \Phi(t)=\Delta
 V-\frac{j(t)}{2}.
\end{equation}

The two-parameter family of self-similar solutions for the parameters
$\Delta \Phi$ and $\varepsilon$ is found in \cite{Dem,KD,DemSh} and now is
compared with the DNS of \eqref{eq1Dim}--\eqref{eq277Dim}. In Fig.~\ref{rho}
(left), such a comparison for the charge distribution
$\rho(\eta){\,=\,c^+(\eta)\,-\,}c^-(\eta)$ is presented for different moments of
time where the parameters of the self-similar solutions are recalculated
according to \eqref{num100}. In Fig.~\ref{rho} (right) universal
voltage--current characteristics $J$ versus $\Delta F=\varepsilon \Delta \Phi$
are shown by a solid line while triangles, stars and circles are gathered for
several fixed drops of potential between the membranes, $\Delta V$, for
intermediate times. The numerical points for all $\Delta V$ shrink into the
universal self-similar VC curve. A rather good correspondence justifies a
self-similar behavior at intermediate times.

\begin{figure}[hbtp]
\begin{center}
    \includegraphics[width=7.85cm]{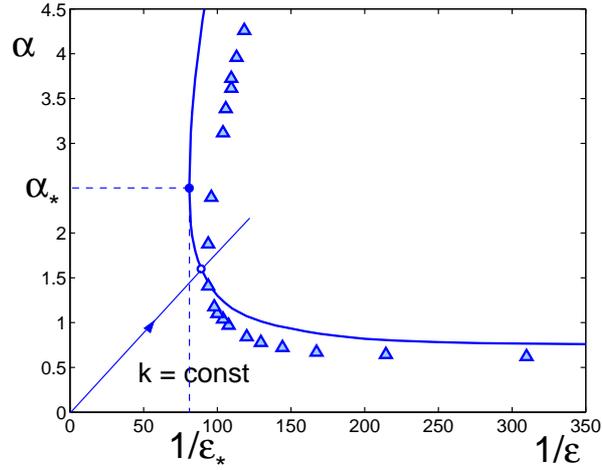}
\end{center}
\caption{Marginal stability curve at $\Delta V=40$ and $\nu=0.001$. The
time-dependent coordinates are inverse Debye length $1/\varepsilon$ and wave
number $\alpha$. Each straight line corresponds to a perturbation with a given
wave number $k$; a point on the line is moving with time towards the stability
curve and eventually crosses it.}\label{rho3}
\end{figure}

\begin{figure}[hbtp]
\begin{center}
    \includegraphics[width=10cm]{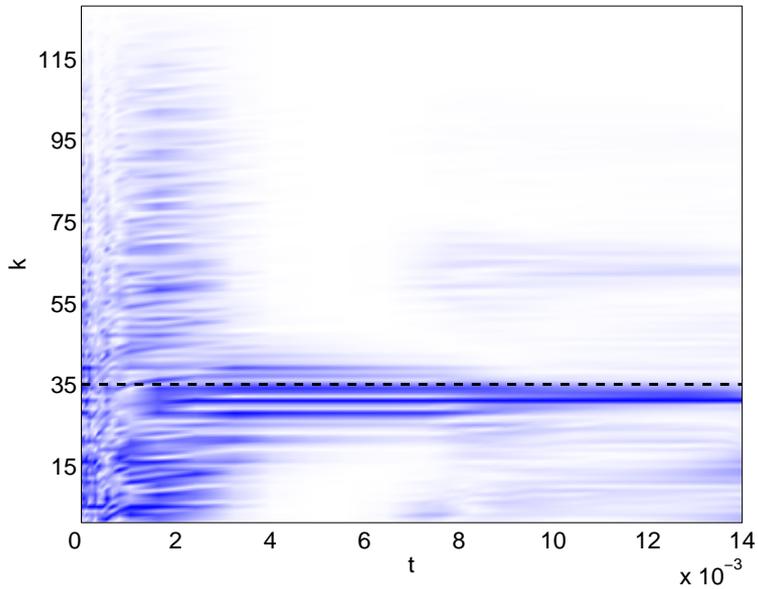}
\end{center}
\caption{Wave number selection for the room disturbances; darker regions
correspond to larger amplitudes. Broadband initial noise is transformed into
a sharp band near $k \approx 32 \div 34$, dashed line corresponds to the
prediction of stability theory, $k_*=35$, $\Delta V=50$ and
$\nu=0.001$.}\label{rho4}
\end{figure}

\begin{figure}[hbtp]
\begin{center}
    \includegraphics[angle=270,width=7.85cm]{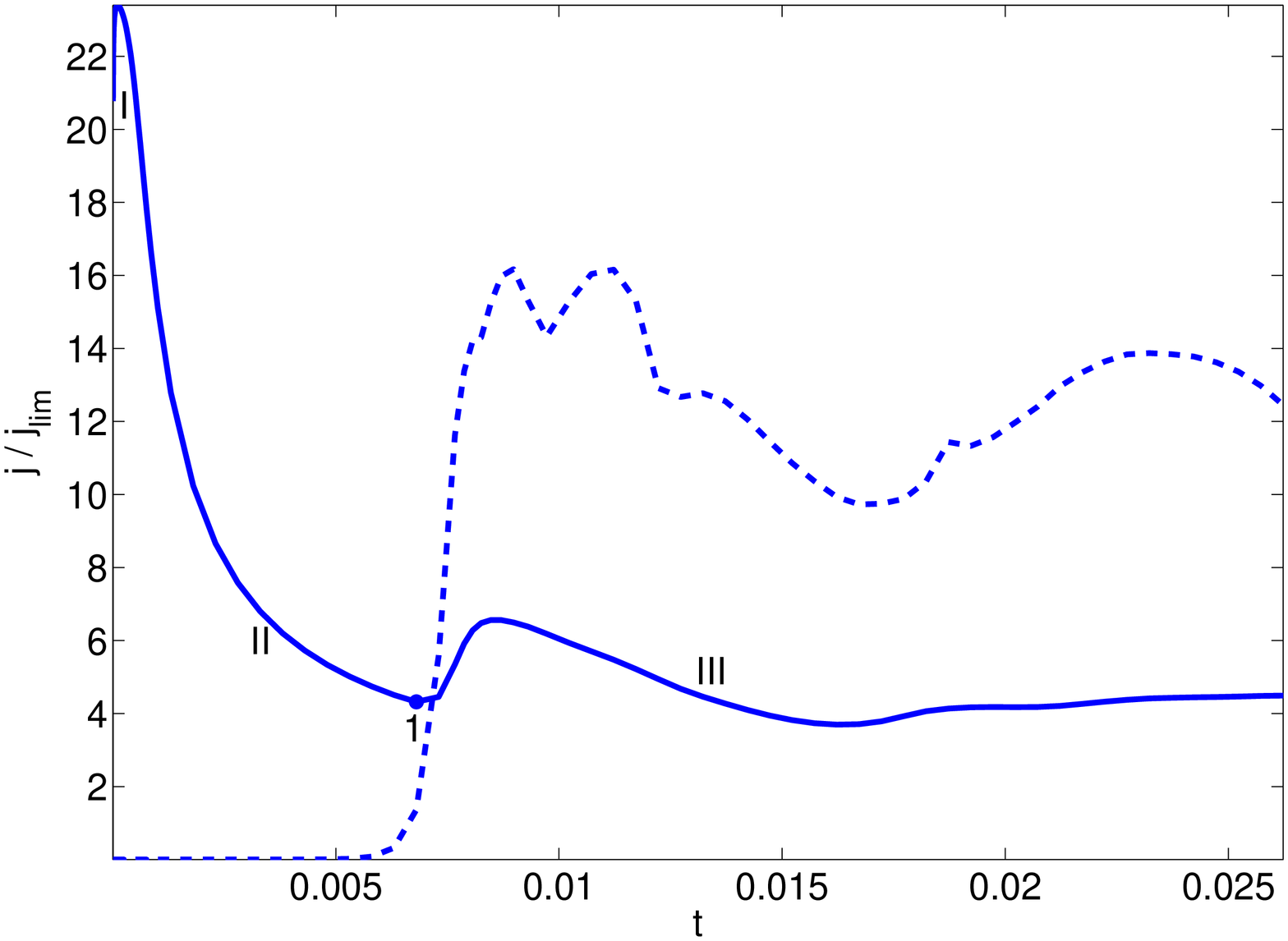}
    \includegraphics[angle=270,width=7.85cm]{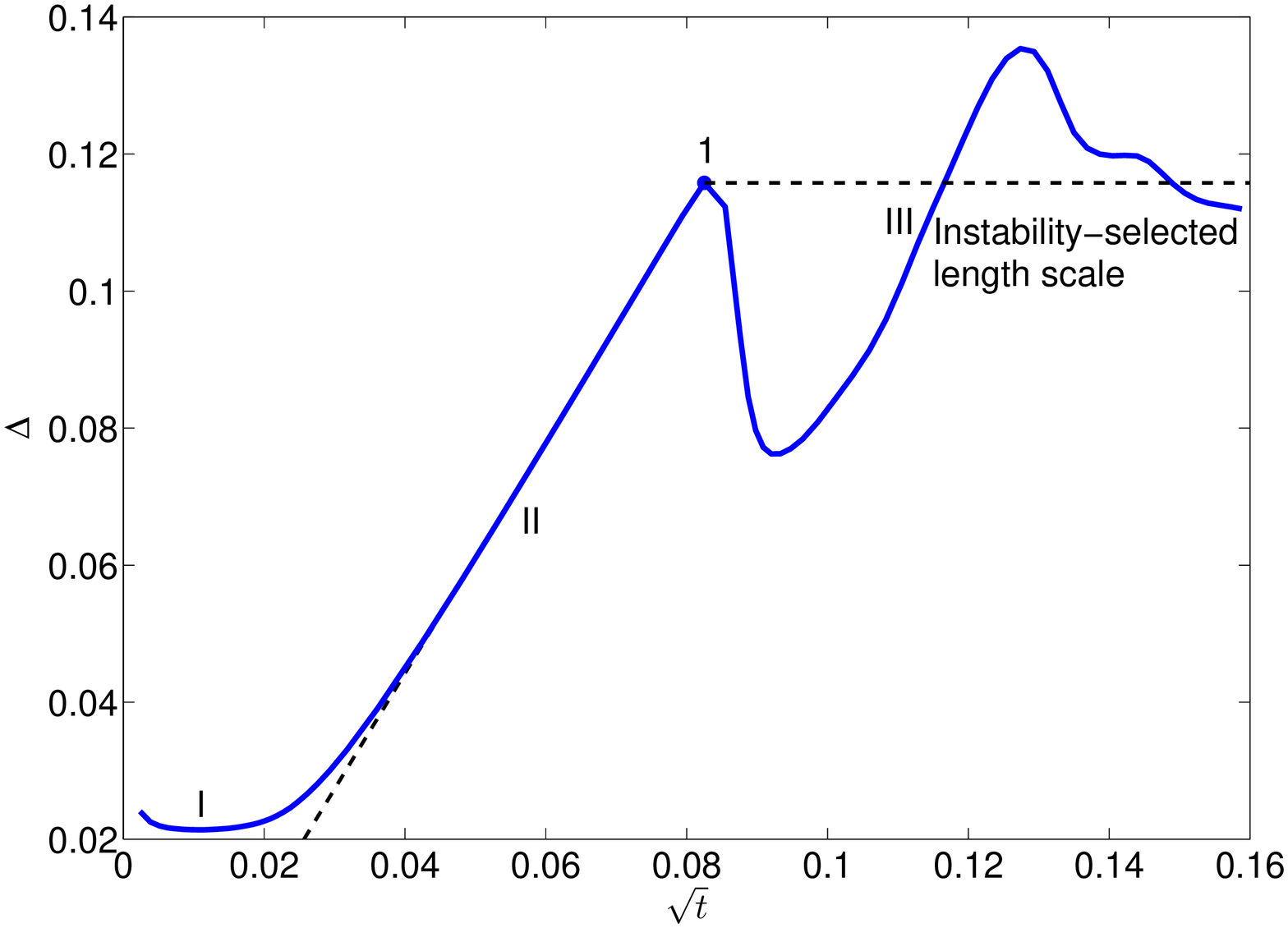}
\end{center}
\caption{Typical evolution of the average electric current $\langle j\rangle
/j_{\lim}$ (solid line) and current amplitude $j_{\max}-j_{\min}$ (dashed line)
(left) and diffusion layer thickness (right) from room disturbances for
$\nu=0.001$ and $\Delta V=60$. Three time intervals are exhibited; point 1
specifies the instability selected length scale.}\label{VACM}
\end{figure}

\header{Instability of the self-similar solution and the length scale
selection.}
The initial stage of evolution is one-dimensional and self-similar, but room
disturbances \eqref{eq277Dim} for the super-critical case,
$\Delta V > \Delta V_*$, will grow and eventually manifest themselves,
destroying the self-similarity and making the solution two-dimensional. This
specifies the next step of evolution. When the amplitudes of the harmonics are
small, their behavior can be considered independently. Imposing on the 1D
self-similar solution a perturbation of the form $\hat{R}(\eta,t)
\exp{(i\alpha x)}$ and linearizing \eqref{eq1Dim}--\eqref{eq277Dim} with respect
to $\hat{R}$ turns this system into a linear PDE system with time-dependent
coefficients. Restricting ourselves to neutral perturbations,
$\partial/\partial t=0$, transforms this PDE system into an ODE system in
$\eta$. Solution of the eigenvalue problem for this system gives the marginal
stability conditions \cite{Dem}.

The parametric dependence of the Debye length and wave number for the
self-similar solution in time completely changes the interpretation of the
marginal stability curves \cite{DemSh}. It is instructive to present the
stability curve at a fixed drop of potential $\Delta V$ in coordinates of the
inverse Debye length $1/\varepsilon=2\sqrt{t}/\nu$ versus the wave number
$\alpha=k \nu/\varepsilon=2k\sqrt{t}$, both referred to the self-similar length
$\sqrt{4 \tilde{D} \tilde{t}}$ and, so, time-dependent, see Fig.~\ref{rho3}.
Each straight line in the figure corresponds to a monochromatic wave with a
fixed wave number, $k=\tilde{\alpha}\tilde{L}$; the larger is $k$, the steeper
is the line. Inside the marginal stability curve a perturbation is growing, but
outside --- decaying. A point on the straight line corresponding to some initial
perturbation is moving with time towards the marginal stability curve just
because of the parametrical time dependence of axes. At the same time, the
amplitude of the disturbance at the beginning is always decaying but after
crossing the marginal curve at some point, it starts to grow. The nose critical
point with coordinates $(1/\varepsilon_*, \alpha_*)$ first reaches the
marginal stability curve and, hence, first loses its stability.
Triangles in the figure are taken from our DNS for monochromatic perturbations.
There is a rather good quantitative agreement between the numerical simulations
and the stability theory for self-similar solutions.

Room disturbances  at the linear stage of evolution can be considered as a
superposition of individual harmonics. Their evolution always starts in the
stable region and filters a broadband initial noise \eqref{eq277Dim} into a
sharp band of wave numbers. It stands to reason that we may assume that the wave
number selection is determined by the critical points $(1/\varepsilon_*,
\alpha_*)$ with the characteristic wave number $k_*=\alpha_*\varepsilon_*/\nu$.
This principle is different from the well-known selecting principle of maximum
growing mode and is verified by comparison of the results of the DNS with the
BC's \eqref{eq277Dim}: the comparison for typical conditions presented in
Fig.~\ref{rho4} shows the validity of this novel principle.

The relation $\Delta(t)=2/\langle j(t)\rangle$ between the diffusion layer
thickness $\Delta$ and the electric current averaged along the membrane length
$l$, $\langle j\rangle=\frac{1}{l}\int_{0}^{l}j(x,t)dx$, is accepted. For the
self-similar solution the diffusion layer thickness should be proportional to
the square root of time, $\Delta(t)\sim\sqrt{t}$, and, respectively,
$\langle j(t)\rangle\sim 1/\sqrt(t)$. The typical evolution of the electric
current $\langle j\rangle$ and of the diffusion layer thickness $\Delta$,
presented in Fig.~\ref{VACM}, exhibit three characteristic time intervals. In
the interval~II, for times of intermediate magnitude, $\Delta$ is proportional
to $\sqrt{t}$ and the behavior is one-dimensional and self-similar, while for
shorter times, in the time interval~I, as well as for longer times in the
interval~III, this self-similarity is violated. In region~I, the influence of
the initial conditions is important, while in region~III, the growth of $\Delta$
is arrested by instability. The current amplitude $j_{\max}-j_{\min}$
characterizes instability and is shown by the dashed line in Fig.~\ref{VACM}
(left); it becomes significant at the point~$1$ which separates regions II and
III and specifies the characteristic length of the diffusion layer. Further
calculations show that this length selected by instability is not changed much
by subsequent nonlinear processes.

\begin{figure}[hbtp]
\begin{center}
    \includegraphics[height=8.5cm]{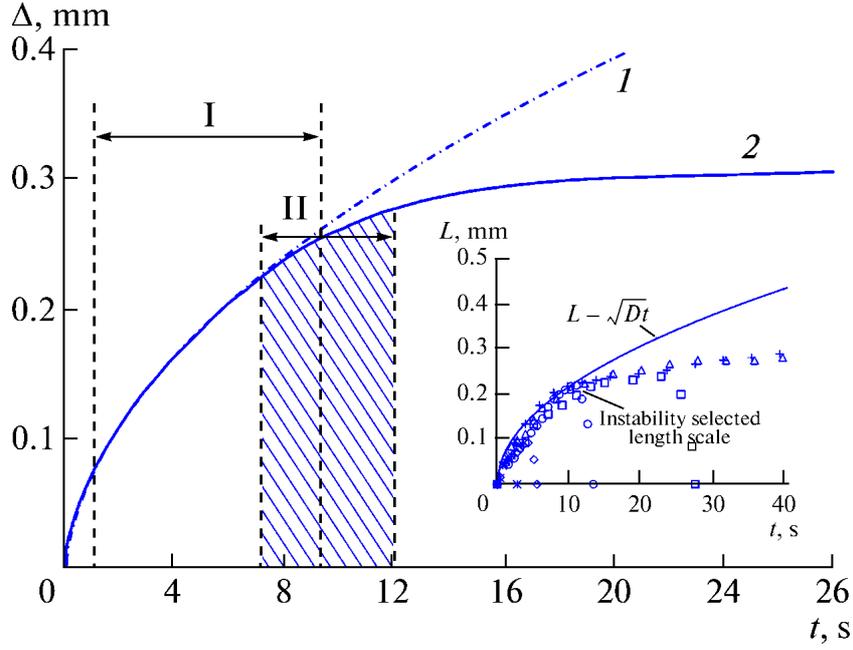}
\end{center}
\caption{Time dependence of the thickness of the diffusion layer
$\tilde{\Delta}$ in dimensional variables for $\Delta \tilde{V} = 2.5$~V,
$\tilde{L}=1$~mm, $\tilde{D}=5\cdot 10^{-9}$~$\mbox{m}^2/\mbox{s}$ and
$\nu=0.001$ \cite{KD}. (1) Self-similar solution, (2) numerical solution, I~is
the self-similarity range, and II~is the stability loss by the trivial solution
and the appearance of electroconvection. The corresponding experimental
dependence \cite{YCh} is in the inset.}\label{invJ}
\end{figure}

In  Fig.~\ref{invJ}, a qualitative comparison with the experiments of Yossifon
and Chang \cite{YCh} is presented (for details see \cite{KD}). The time
dependence of the thickness of the diffusion layer $\tilde{\Delta}$ is taken in
dimensional variables. Excluding very short times of establishment, up to
$8.9$~s of evolution, the solution in zone I follows the self-similar law
proportionally to $\sqrt{\tilde{t}}$ (dash-and-dot line); then the self-similar
process is violated by the loss of stability in zone II. According to the
numerical experiment, the transition to the electroconvection mode takes place
in the range $6 \div 10$~s. This corresponds to the value of 8~s observed in
physical experiments \cite{YCh}.

\header{Nonlinear stages of evolution and secondary instability.}
The next stages of evolution are nonlinear processes with eventual saturation of
the disturbance amplitude. For a small super-criticality, $\Delta V-\Delta V_*$,
the nonlinear saturation leads to steady periodic electro-convective vortices
along the membrane surface; with increasing super-criticality the attractor can
be described as a structure of periodically oscillating vortices;  with further
increase in super-criticality, the behavior eventually becomes chaotic in time
and space.

\begin{figure}[hbtp]
  \begin{center}
    \includegraphics[width=7.85cm]{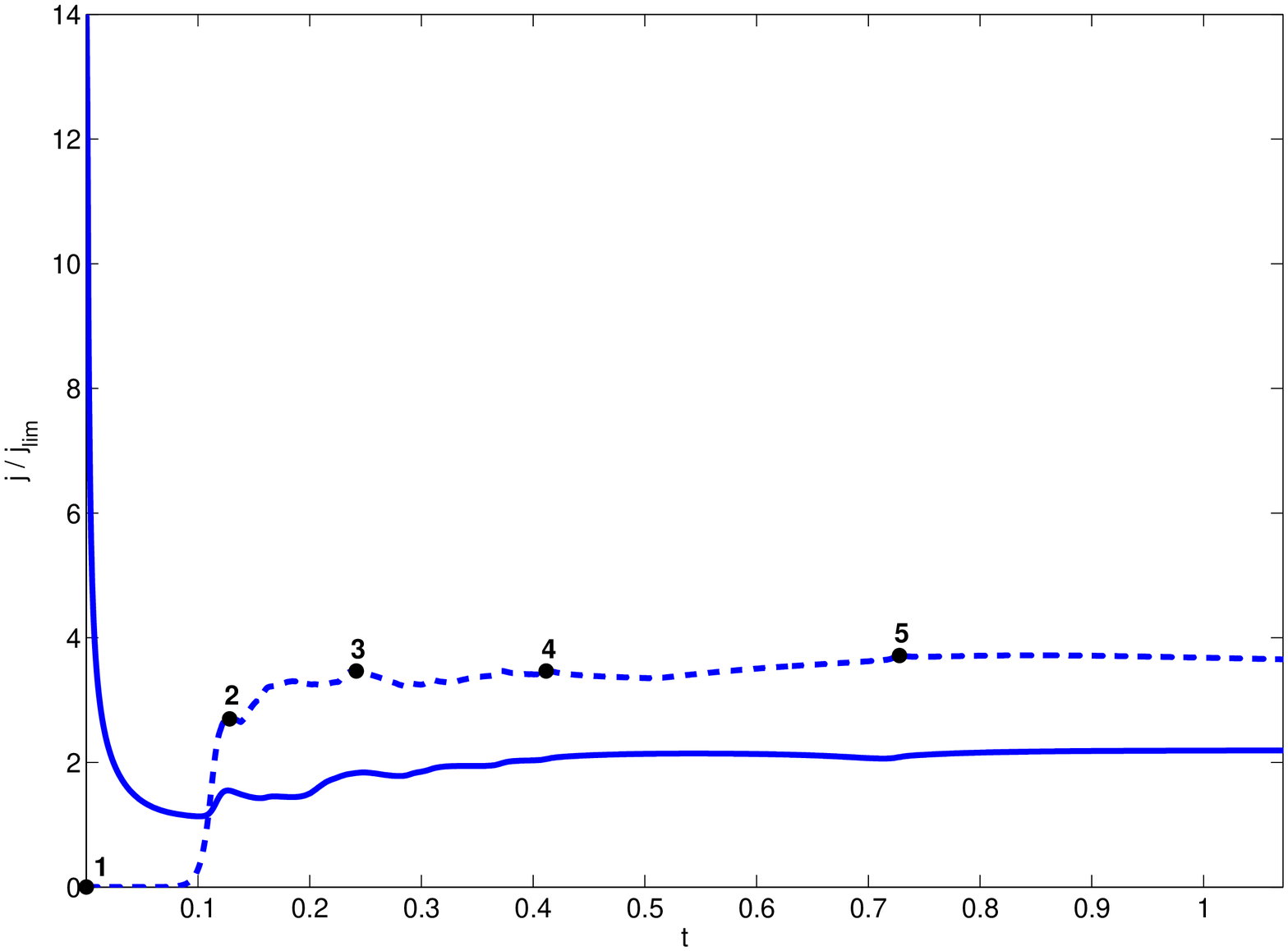}
    \includegraphics[width=7.85cm]{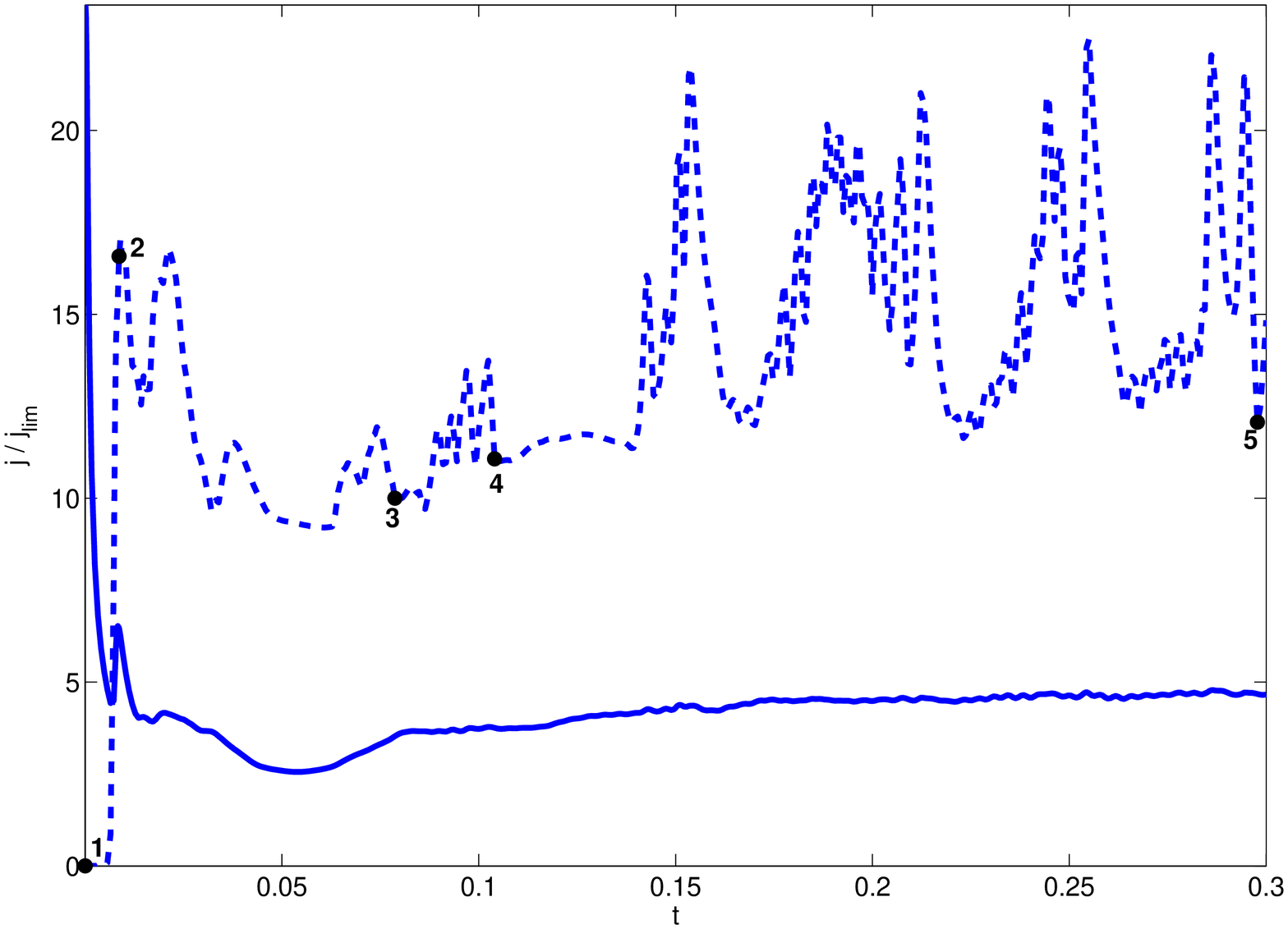}
\end{center}
\caption{Evolution of the normalized average electric current
$\langle j\rangle/j_{\lim}$ (solid line) and current amplitude
$(j_{\max}-j_{\min})/j_{\lim}$\ (dashed line) for $\nu=0.001$
$\kappa=0.2$ for small super-criticality, $\Delta V=30$ (left) and
for large super-criticality, $\Delta V=50$\ (right).}\label{P3p}
\end{figure}

The evolution of the normalized to $j_{\lim}$ average electric current and
current amplitude of perturbation are presented in Fig.~\ref{P3p}. For small
super-criticality this evolution results in steady vortices, Fig.~\ref{P3p}
(left),  and for larger super-criticality --- in chaotic behavior,
Fig.~\ref{P3p} (right).

\begin{figure}[hbtp]
  \begin{center}
   \includegraphics[width=7.85cm]{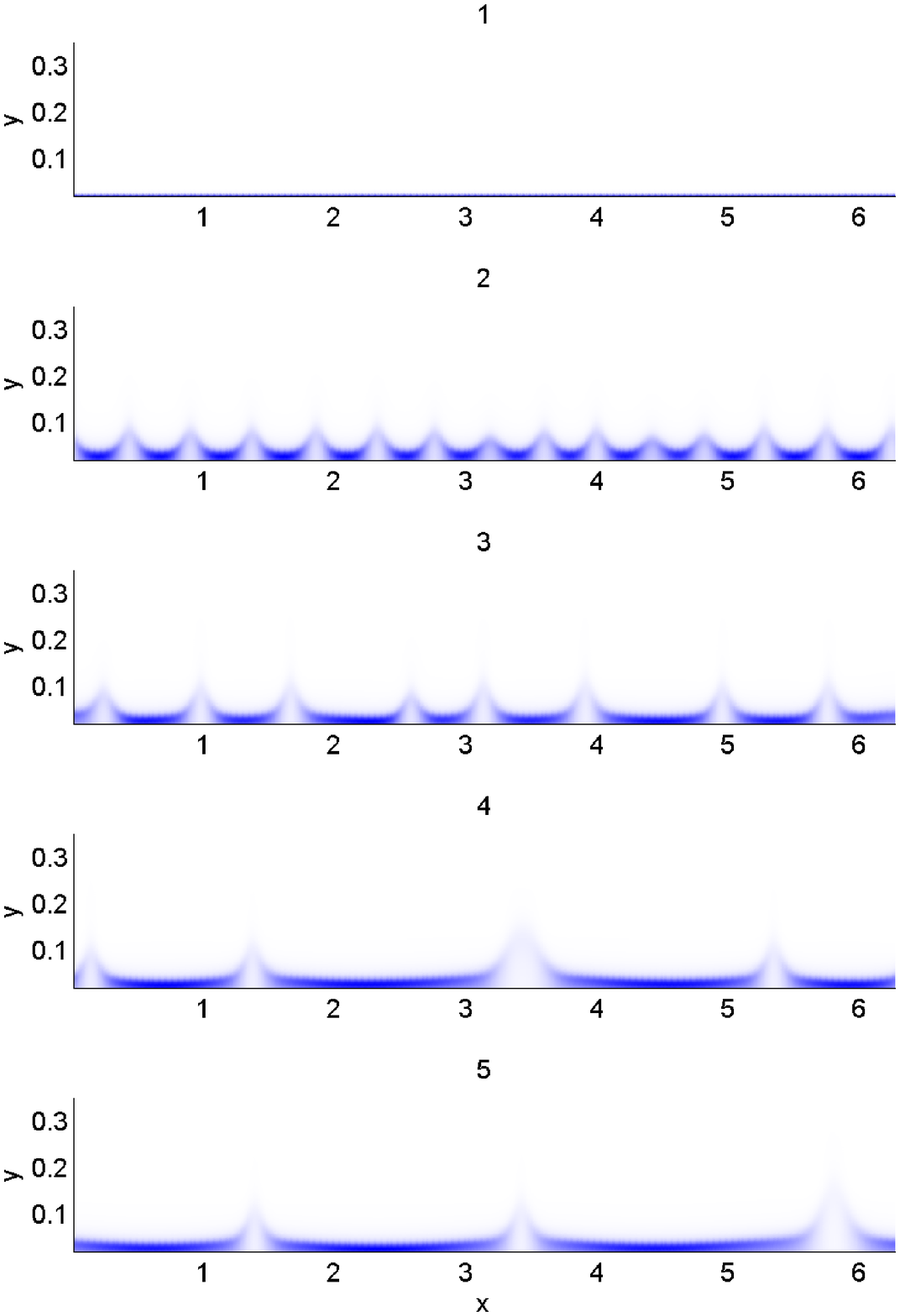}
   \includegraphics[width=7.85cm]{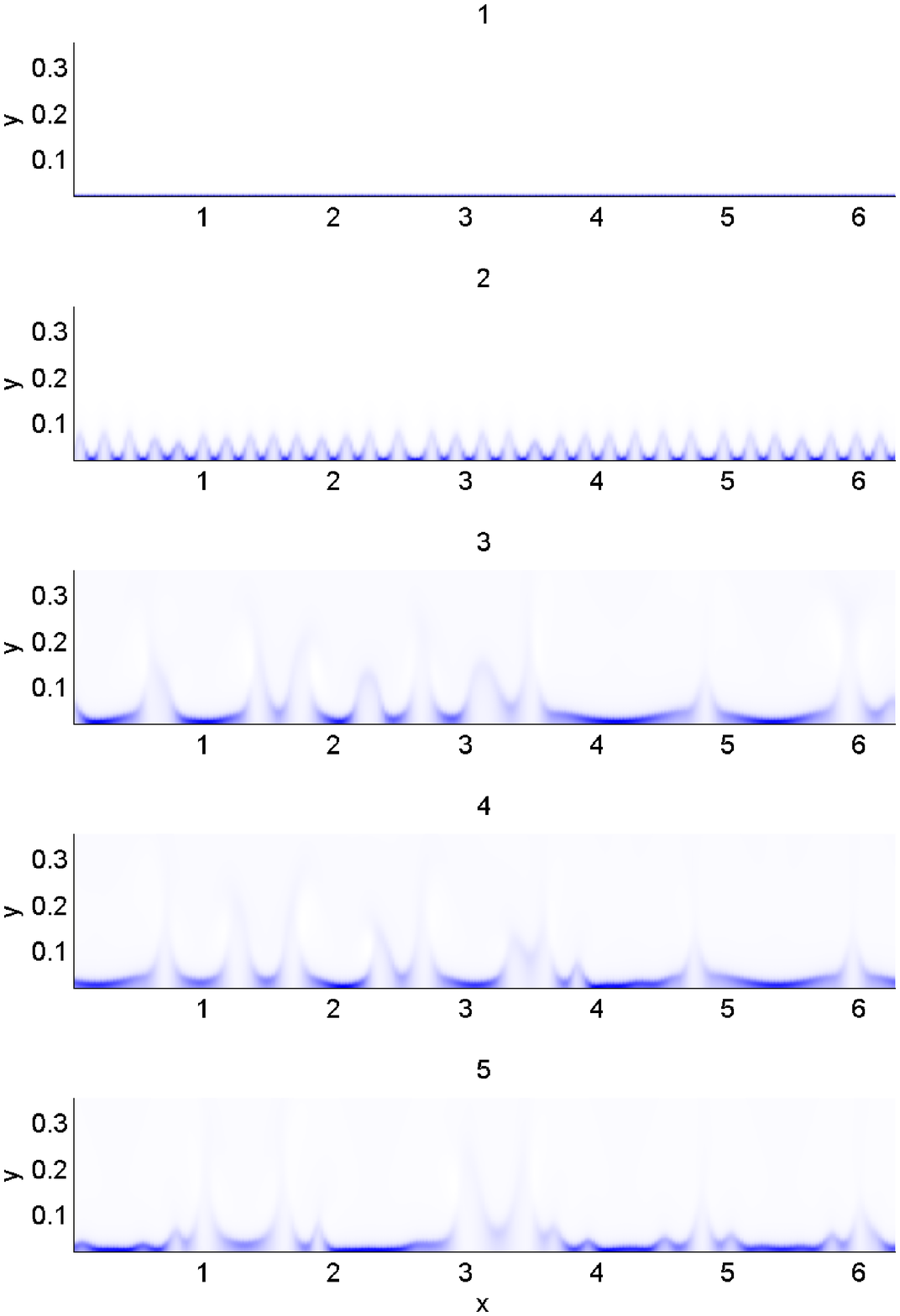}
\end{center}
\caption{Snapshots of the charge density $\rho(x,y,t)$ at
different times. Darker regions correspond to larger charge
densities (for parameters, see previous figure). }\label{PAJ}
\end{figure}

Snapshots of the charge density $\rho(x,y,t)$ for the above mentioned conditions
at different times are shown in Fig.~\ref{PAJ}. Darker regions correspond to
larger densities of space charge. There is a rather sharp boundary between the
ESC region and the diffusion region. Point 1 corresponds to the initial
small-amplitude broadbanded white noise. A linear stability mechanism filters it
into practically periodic disturbances with a wave number $k_*$, point 2. The
boundary between the ESC region and the diffusion region behaves sinusoidally:
its minimum corresponds to the maximum of the charge density and vice versa.
With increasing time, there is a fusion of neighboring waves with corresponding
reducing of their number. The sinusoidal profile changes to the spike-like one
where regions of small charge in the spikes are jointed by thin flat regions of
large space charge, moment 5, left. Besides fusion of the spikes there is also
their creation so that eventually some equilibrium of their number is
established, moment 5, right. Movie visualization for charge density evolution
can be found in \cite{CT} ($\Delta V=50$, $\nu=10^{-3}$ and $\kappa=0.2$,
chaotic behavior).

\begin{figure}[hbtp]
\begin{center}
    \includegraphics[width=7.5cm]{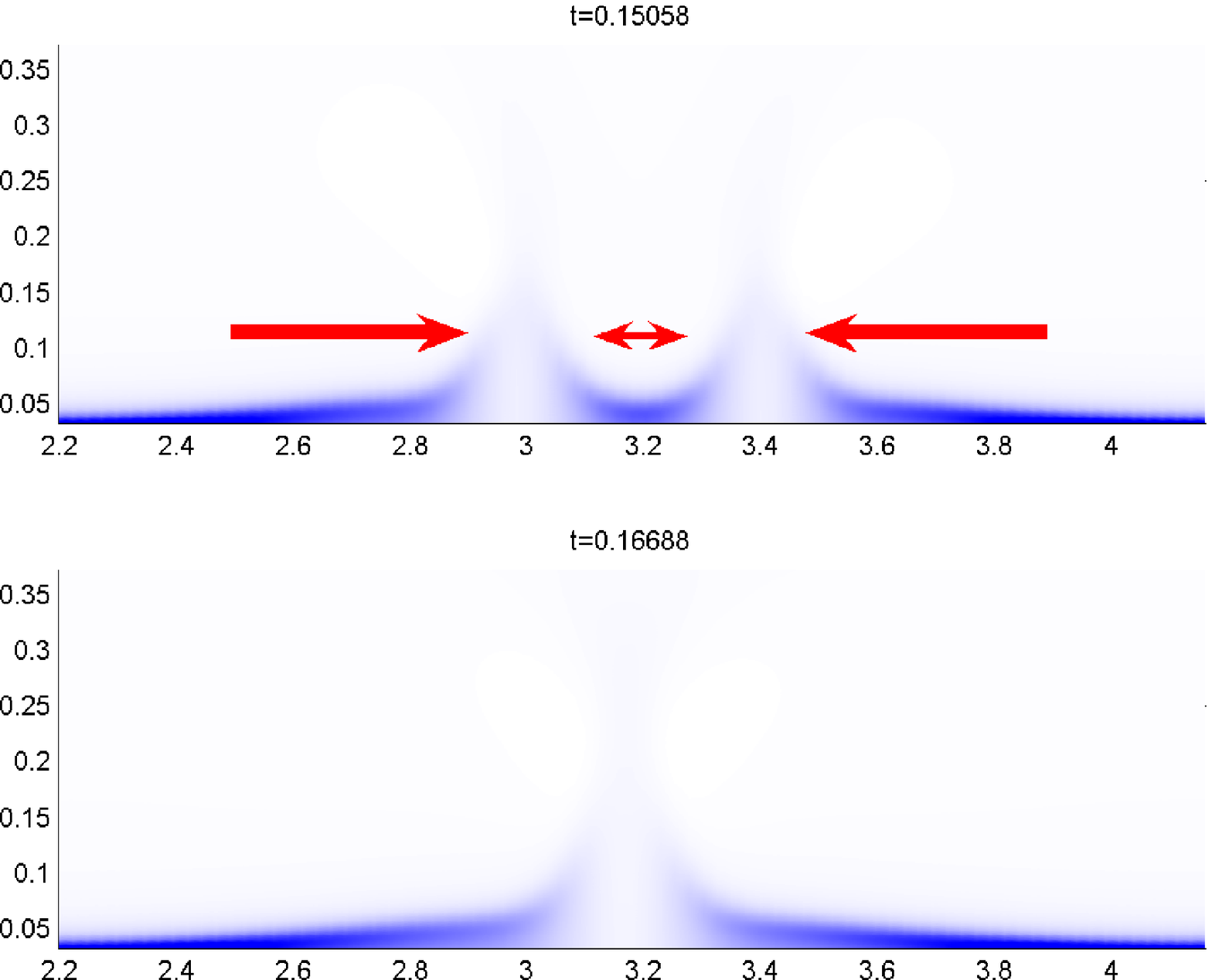}
    \includegraphics[width=7.5cm]{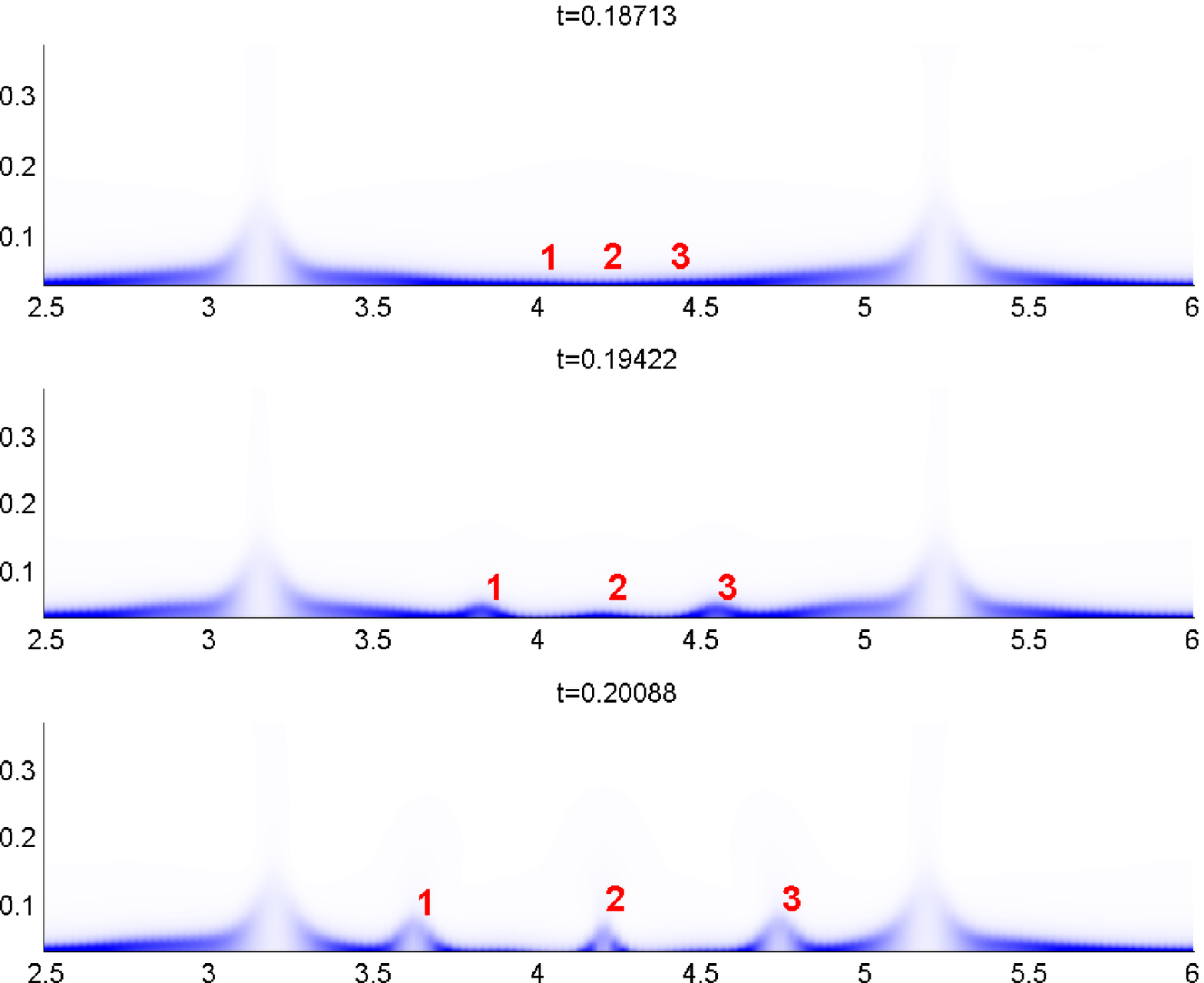}
\end{center}
\caption{Two physical mechanisms of the secondary instability. (a) Coalescence
of two humps with low charge density with following coarsening. (b) Instability
of flat region between humps with following creation of new humps.}\label{Cm}
\end{figure}

Let us consider the physical mechanisms of the secondary instability. The sharp
boundary between the ESC and diffusion regions has small charge densities near
its humps and large charge densities in neighboring regions around its hollows,
either for periodic or spike-like coherent structures. These regions with larger
charge are trying to expand (charge of the same sign is repelling) and this can
lead to the disappearance of some of the humps --- with eventual coarsening.
Such instability is illustrated in Fig.~\ref{Cm} (left). There is also another
physical mechanism which results in the opposite effect. If a flat region
between two humps is long enough, it suffers from primary Rubinstein and
Zaltsman electroconvective instability, see Fig.~\ref{Cm} (right). Points 1, 2
and 3 are nucleation points of future humps. They grow and eventually form new
humps. This last mechanism is applicable only for spike-like structures with
flat neighboring regions, but not for sinusoidal waves.

%%%%%%%%%%%%%%%%%%%%%%%%%%%%%%%%%%%%%%%%%%%%%%%%%%%%%%%%%

\header{Conclusions.}
The results of the DNS reported in this paper refer to  realistic experimental
parameters that may be realized in lab experiments. The following stages of
evolution are identifed:
i) a stage in which the initial conditions have considerable influence
(milliseconds);
ii) 1D self-similar evolution (milliseconds--seconds);
iii) primary instability of the self-similar solution (seconds);
iv) a nonlinear stage  with secondary instabilities. This work can be viewed as
a new step in the understanding of the instabilities and pattern formation in
micro- and nano-scales.

The authors are grateful to I. Rubinstein, M. Z. Bazant, and H.- C. Chang for
fruitful discussions. This work was supported in part by Russian Foundation for
Basic Research (project no. 11-08-00480-a).

\newpage

{\bf Figure Captions}
\begin{itemize}

\item[\bf Figure 1.] (Left) Distribution of the charge density $\rho$ in space
for several moments of time, $\Delta V=100$ and
$\nu=\tilde{\lambda_D}/\tilde{L}=0.001$; (a) --- numerical solution of
\eqref{eq1Dim}--\eqref{eq277Dim}, (b) --- self-similar solution.  The numerical
points for $\nu=0.0005$, and several values of  $\Delta V$ shrink into
one self-similar VC curve (shown by the solid line).

\item[\bf Figure 2.] Marginal stability curve  at  $\Delta V=40$ and
$\nu=0.001$. The time-dependent coordinates are inverse Debye length
$1/\varepsilon$ and wave number $\alpha$. Each straight line corresponds to a
perturbation with a given wave number $k$; a point on the line is moving with
time towards the stability curve and eventually crosses it.

\item[\bf Figure 3.] Wave number selection  for the room disturbances; darker
regions correspond to larger amplitudes. Broadband initial noise is transformed
into a sharp band near $k \approx 32 \div 34$, dashed line corresponds to the
prediction of stability theory, $k_*=35$, $\Delta V=50$ and $\nu=0.001$.

\item[\bf Figure 4.] Typical evolution of the average electric current
$\langle j\rangle/j_{\lim}$ (solid line) and current amplitude
$j_{\max}-j_{\min}$ (dashed line) (left) and diffusion layer thickness (right)
from room disturbances for $\nu=0.001$ and $\Delta V=60$. Three time intervals
are exhibited; point 1 specifies the instability selected length scale.

\item[\bf Figure 5.] Time dependence of the  thickness of the diffusion layer
$\tilde{\Delta}$ in dimensional variables for $\Delta \tilde{V} = 2.5$~V,
$\tilde{L}=1$~mm, $\tilde{D}=5\cdot 10^{-9}$~$\mbox{m}^2/\mbox{s}$ and
$\nu=0.001$ \cite{KD}. (1) Self-similar solution, (2) numerical solution, I~is
the self-similarity range, and II~is the stability loss by the trivial solution
and the appearance of electroconvection. The corresponding experimental
dependence \cite{YCh} is in the inset.

\item[\bf Figure 6.] Evolution of the normalized average electric current
$\langle j\rangle/j_{\lim}$ (solid line) and current amplitude
$(j_{\max}-j_{\min})/j_{\lim}$ (dashed line) for $\nu=0.001$ $\kappa=0.2$ for
small super-criticality, $\Delta V=30$ (left) and for large super-criticality,
$\Delta V=50$\ (right).

\item[\bf Figure 7.] Snapshots of the charge density $\rho(x,y,t)$ at different
times. Darker regions correspond to larger charge densities (for parameters,
see previous figure).

\item[\bf Figure 8.] Two physical mechanisms of the secondary instability. (a)
Coalescence of two humps with low charge density with following coarsening. (b)
Instability of flat region between humps with following creation of new humps.

\end{itemize}

\end{document}